\documentclass[aps,twocolumn,showpacs, superscriptaddress,amsmath,amssymb,nofootinbib]{revtex4-1}
\usepackage{graphicx}
\usepackage{epsfig}
\usepackage{epstopdf}
\usepackage{bm}% bold math
\usepackage[usenames,dvipsnames]{color}
\usepackage{float}

\begin{document}
\title{Valence band electronic structure of the van der Waals ferromagnetic insulators: VI$_3$ and CrI$_3$}
\author{Asish K. Kundu}
\email{akundu@bnl.gov}
%\author{Yu Liu\textsuperscript{$\ddag,$}}
\author{Yu Liu}
\email{present address: Los Alamos National Laboratory, MS K764, Los Alamos NM 87545}
\author{C. Petrovic}
\author{T. Valla}
\email{valla@bnl.gov}
\affiliation{Condensed Matter Physics and Materials Science Department, Brookhaven National Laboratory, Upton, New York 11973, USA}
%\let\thefootnote\relax\footnote{$^{\ddag}$ Present address: Los Alamos National Laboratory, MS K764, Los Alamos NM 87545}
%\date{\today}

\begin{abstract}
Ferromagnetic van der Waals (vdW) insulators are of great scientific interest for their promising applications in spintronics. It has been indicated that in the two materials within this class, CrI$_3$ and VI$_3$, the magnetic ground state, the band gap, and the Fermi level could be manipulated by varying the layer thickness, strain or doping. To understand how these factors impact the properties, a detailed understanding of the electronic structure would be required. However, the experimental studies of the electronic structure of these materials are still very sparse. Here, we present the detailed electronic structure of CrI$_3$ and VI$_3$ measured by angle-resolved photoemission spectroscopy (ARPES). Our results show a band-gap of the order of 1 eV, sharply contrasting some theoretical predictions such as Dirac half-metallicity and metallic phases, indicating that the intra-atomic interaction parameter (U) and spin-orbit coupling (SOC) were not properly accounted for in the calculations. We also find significant differences in the electronic properties of these two materials, in spite of similarities in their crystal structure. In CrI$_3$, the valence band maximum is dominated by the I 5{\it p}, whereas in VI$_3$ it is dominated by the V 3{\it d} derived states. Our results represent valuable input for further improvements in the theoretical modeling of these systems.
\end{abstract}

\maketitle

\section*{Introduction}

Since the discovery of graphene \cite{geim2010rise} and transition-metal dichalcogenides \cite{wang2012electronics}, the quest for magnetic van der Waals materials is under way because of  possible applications in spintronic devices \cite{mcguire2017magnetic,huang2017layer,seyler2018ligand,gong2017discovery,ajayan2016van,wang2018very,Otrokov2019,Gui2019}. Recently, the intense theoretical and experimental studies are focused on Cr$_2$Ge$_2$Te$_6$ \cite{liu2017critical,suzuki2019coulomb,gong2017discovery,jiang2020spin,liu2019anisotropic,li2018electronic}, Cr$_2$Si$_2$Te$_6$ \cite{liu2019anisotropic}, CrI$_3$ \cite{liu2019thickness,huang2017layer,wang2018very,seyler2018ligand,webster2018distinct,liu2018anisotropic,lado2017origin,lee2020fundamental,liu2018three} and VI$_3$ \cite{yan2019anisotropic,tian2019ferromagnetic,huang2020discovery,son2019bulk,kong2019vi3,liu2020critical,long2020stacking,he2016unusual,yang2020vi}. These materials show an interesting development of magnetic properties, going from bulk, down to a single layer \cite{huang2017layer,seyler2018ligand}. Theoretically, it has been demonstrated that a strain tuning can result in an enhancement of magnetic ordering temperature, or in switching of magnetic state from the ferromagnetic (FM) to an antiferromagnetic (AFM) half-semiconductor in CrI$_3$ and VI$_3$ monolayers and in their heterostructures \cite{wu2019strain,subhan2020magnetic,subhan2020large,jiang2018controlling}. Also, the Dirac half-metallicity and the twin orbital-order phases were predicted to exist in a monolayer of VI$_3$ \cite{he2016unusual,huang2020discovery}.

Furthermore, some truly remarkable phenomena have been recently investigated in the CrI$_3$ bilayer where the AFM coupling between the two FM layers simultaneously breaks both the time-reversal ($T$) and the inversion ($P$) symmetries in an otherwise centrosymmetric crystal. This symmetry breaking leads to a giant nonreciprocal second-harmonic generation \cite{Sun2019} and a magnetic photogalvanic effect \cite{zhang2019switchable}, or creation of dc current by a linearly polarised light, making these materials promising candidates for the future applications in which the coupling between the magnetic, optical and electronic degrees of freedom can be utilized to manipulate the properties. Also, heterostructures involving the vdW ferromagnets and superconductors could result in the elusive spin triplet superconductivity and could serve as extremely effective spin valves \cite{soriano2020magnetic}.

Both CrI$_3$ and VI$_3$ belong to a family of transition metal trihalides MX$_3$ (M=Cr, V and X = Cl, Br, and I) \cite{dillon1965magnetization} where Cr and V ions are octahedrally coordinated by halide ions, forming a honeycomb lattice within the a-b plane. In case of CrI$_3$, the Cr$^{3+}$ has half filled t$_{2g}$ level yielding S = 3/2 \cite{lee2020fundamental}, while VI$_3$, V$^{3+}$ has two valence electrons in the t$_{2g}$ states, yielding S = 1 \cite{son2019bulk}. Bulk CrI$_3$ and VI$_3$ are layered insulating 2D ferromagnets with $T_C$ = 61 K and 50 K \cite{liu2018anisotropic,liu2020critical} and band gap of 1.24 eV and 0.6-0.67 eV, respectively \cite{dillon1965magnetization,son2019bulk}. Upon cooling, CrI$_3$ undergoes a phase transition $\sim$~220 K from the high-temperature monoclinic (C2$/$m) to the low-temperature rhombohedral (R{$\bar3$}) phase \cite{mcguire2015coupling}. VI$_3$ also undergoes a structural phase transition at $\sim$~80 K. Above transition temperature it has a trigonal structure (space group R${\bar3}$) \cite{liu2020critical,kong2019vi3} but exact symmetry of the low-temperature structure is not established yet \cite{son2019bulk,tian2019ferromagnetic,wang2020raman}. The experimental studies on these systems are mainly focused on the magnetic and structural properties \cite{liu2018anisotropic,wang2018very,tian2019ferromagnetic,son2019bulk,kong2019vi3,yan2019anisotropic,liu2020critical,dolevzal2019crystal}. Magnetic measurements show the presence of magnetocrystalline anisotropy (MCA) in these systems, favoring an out-of-plane orientation of magnetic moments \cite{liu2018anisotropic,yan2019anisotropic,son2019bulk,subhan2020magnetic}. It is known that MCA in these systems originates due to the strong SOC induced by the iodine ligand \cite{lee2020fundamental,lado2017origin,yang2020vi}.

The experimental studies of the electronic structure in this class of materials were restricted to Cr$_2$Ge$_2$Te$_6$ and Cr$_2$Si$_2$Te$_6$ and to the best of our knowledge, no reports are available on CrI$_3$ and VI$_3$ \cite{suzuki2019coulomb,li2018electronic,jiang2020spin}. Even though the magnetic and optical measurements indicate that the strong SOC and U play an important role, the band structure calculations often do not take into account these parameters, erroneously predicting properties such as Dirac half-metallicity or metallic phases \cite{he2016unusual,baskurt2020vanadium,webster2018distinct}. The calculations that do account for U predict Mott-insulating ground state, both in monolayer as well as bulk VI$_3$, in agreement with the experiments \cite{an2019tuning,tian2019ferromagnetic,son2019bulk}.

Webster {\it et al.} \cite{webster2018distinct} have calculated the band structure of a monolayer CrI$_3$ for non-magnetic, FM and AFM phases using local density approximation (LDA) and generalized-gradient approximation (GGA). These calculations show that the introduction of magnetism   leads to metal-to-semiconductor transition. In addition, they indicate that the nature of band gaps (direct or indirect) in magnetic phases shows a delicate dependence on the magnetic ordering and spin-orbit coupling. Zhang {\it et al.} \cite{zhang2015robust} have performed calculations for bulk and monolayer CrX$_3$ (X=F, Cl, Br and I) using two different methods. Their results show that a monolayer and bulk have very similar partial- and total DOS but the relative contributions within the DOS and energy positions differ significantly between the two methods. Orbital resolved band dispersions for a monolayer CrI$_3$ is reported by Jiang {\it et al.} \cite{jiang2018spin} and Kim {\it et al.} \cite{kim2019exploitable}. They show that the the valence band maximum (VBM) is dominated by I 5$p$ orbitals (mainly $p_x$ and $p_y$) and have also discussed the effects of SOC in the band dispersions. The density functional theory (DFT) calculations \cite{lado2017origin} using DFT+U also show the dominance of I 5$p$ orbitals in the VBM of CrI$_3$ and highlight the importance of SOC for the observed magnetic anisotropy. Moreover, depending on the calculation method and choice of parameters, the resulting electronic structure of the same material may differ significantly, leaving huge discrepancies in predicted material\rq{}s properties \cite{zhang2015robust,jiang2018spin,gudelli2019magnetism,wang2020electronic,an2019tuning}. These uncertainties emphasize the importance of experimental studies of the electronic structure of these materials.

Here, we present the first experimental studies of the valence band electronic structure of CrI$_3$ and VI$_3$ using ARPES. Even though both materials have a similar crystal structure, our results indicate significant differences in their electronic structure. In CrI$_3$ we find that the VBM is dominated by I 5{\it p} contribution, while in VI$_3$ the VBM is mainly of V 3{\it d} character. A comparison of the experimental band dispersions with the theoretical ones indicates that even the monolayer band structure captures most of the features seen experimentally. Our results illustrate the importance of SOC and U in these materials, ruling out various metallic phases predicted by theories in which these parameters were neglected. Our study clearly shows the semiconducting/insulating behavior, with the band gap in the ~1 eV range, and provides useful input for the improvements in the theoretical understanding of these systems.

\section*{Results and discussions}

\begin{figure*}[ht!]
\centering
\includegraphics[width=17cm]{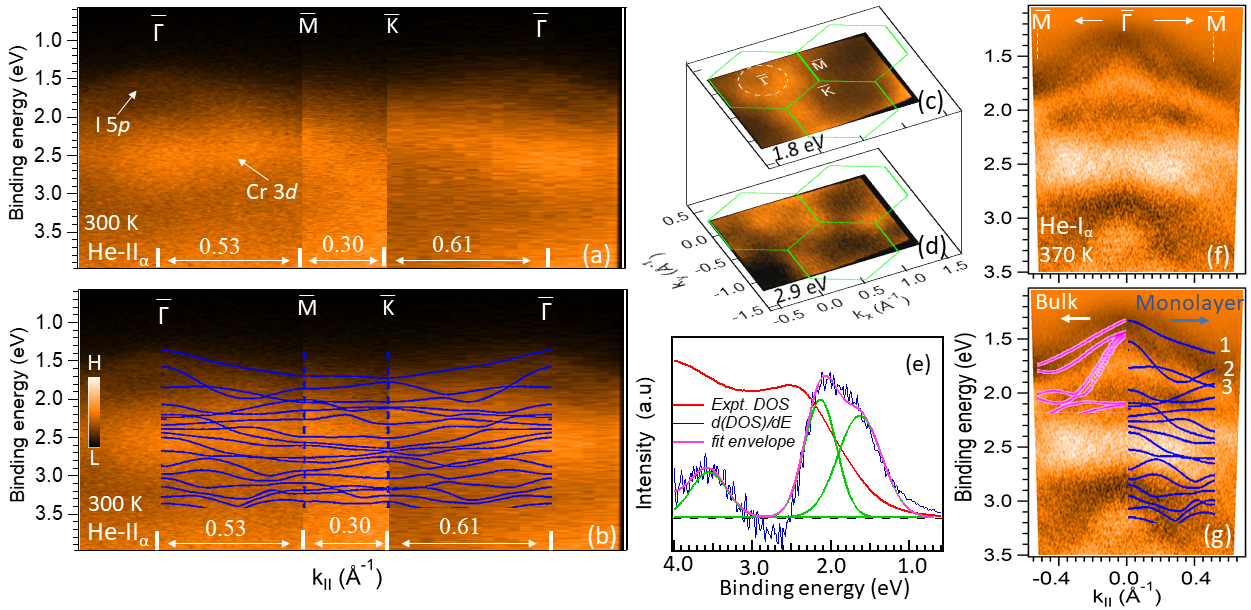}
\caption {Valence band electronic structure of CrI$_3$. (a) Experimental electronic structure along the $\bar{\Gamma}$-$\bar{M}$-$\bar{K}$-$\bar{\Gamma}$ path using He-II$\alpha$ photons at 300 K. (b) Theoretical band structure for a CrI$_3$ monolayer \cite{jiang2018spin}, superimposed on the measured one from (a). (c) and (d) ARPES iso-energy contours at two different energies, 1.8 and 2.9 eV from the Fermi level, respectively. Dotted-circle represents the shape of the contour around $\bar{\Gamma}$. (e) Density of states (DOS) obtained by integration of spectral intensity over the whole Brillouin zone and its differentiation ($d(DOS)/dE$). To resolve the peak positions, differentiated DOS (blue) is fitted with three gaussian peaks (green). Peaks at 1.6 eV and 2.2 eV are dominated by I $5p$ and Cr $3d$ states, respectively. (f) Second derivative of band dispersions ($d^2I/dE^2$) along $\bar{\Gamma}$-$\bar{M}$ using He-I$\alpha$ photons (21.2 eV) measured at 370 K. (g) Calculated band dispersions with FM configuration and SOC for the monolayer  and bulk \cite{webster2018distinct} CrI$_3$ are superimposed on (f). Bands marked by 1, 2 and 3 are dominated by I 5{\it p} character around $\bar{\Gamma}$.}\label{Fig1}
\end{figure*}

Figure~\ref{Fig1} shows electronic band dispersions of CrI$_3$ at room temperature (300 K) along the high symmetry directions of surface Brillouin zone (SBZ). Band dispersions along the $\bar{\Gamma}$-$\bar{M}$-$\bar{K}$-$\bar{\Gamma}$ path is shown in Fig.~\ref{Fig1}(a). To understand the orbital character of these bands, the calculated band structure of a CrI$_3$ monolayer \cite{jiang2018spin} considering FM ground state and SOC is shown on top of the experimental one in Fig.~\ref{Fig1}(b). The details of orbital character can be found elsewhere in the supplimentary material of ref. \cite{jiang2018spin}. The experimental VBM and the theoretical one were aligned to be at the same energy. Even though our data are from the cleaved bulk crystal, a qualitative comparison with a monolayer should be appropriate because the layers in the crystal are weakly coupled and their electronic structure does not depend significantly on the thickness \cite{zhang2015robust}. Indeed, the overall features in the experimental electronic structure moderately agree with the theory. While the experimental bands are not sharp enough to identify all the fine features existing in the calculations, the measured band dispersions generally agree with the calculations \cite{jiang2018spin}. Some discrepancies may arise due to the absence of U in the calculations. The ARPES spectra are intrinsically broad due to the existence of a large number of overlapping bands \cite{zhang2015robust} and a relatively high temperature of the measurements. The broad, less dispersive features at $\sim$~2.5 eV are mainly of Cr 3{\it d} character, according to calculations \cite{kim2019exploitable,jiang2018spin}. Around the $\bar{\Gamma}$ point, within the energy range 1.3-2.1 eV, a strongly dispersive feature is attributed to iodine 5{\it p} orbital \cite{kim2019exploitable,jiang2018spin}.  In iso-energy contour, this band forms a hole-like circular feature, centered at $\bar{\Gamma}$, as shown in Fig.~\ref{Fig1}(c). This feature resembles the theoretical Fermi surface of the hole-doped CrI$_3$ monolayer with an out-of-plane magnetization \cite{jiang2018spin}. Very similar features are reported for Cr$_2$Ge$_2$Te$_6$ \cite{suzuki2019coulomb,suzuki2019coulomb,jiang2020spin}, a material in the same class, which has predominantly the Te 5{\it p} character at the VBM. The iso-energy contour at 2.9 eV binding energy shows hexagonal symmetry, representing the symmetry of the CrI$_3$ surface (Fig.~\ref{Fig1}(d)).

Figure~\ref{Fig1}(f) shows the second derivative of the ARPES intensity ($d^2I/dE^2$) along the $\bar{\Gamma}$-$\bar{M}$, measured using the He-I$\alpha$ photons. The data were acquired at even higher temperature, 370 K, to minimize the charging. Similar to Fig.~\ref{Fig1}(a), we see the strong bands at $\sim$~2.5 eV originating from the Cr 3{\it d}. In addition, above those, the three more bands can now be identified more clearly. These three bands are mainly of I 5{\it p} character \cite{jiang2018spin,webster2018distinct,lado2017origin,kim2019exploitable} around $\bar{\Gamma}$ and agree moderately with the bands calculated using the generalized gradient approximation (GGA) and accounting for the SOC and the ferromagnetic ground state with the out-of-plane moments \cite{jiang2018spin,webster2018distinct}. The calculated bands \cite{jiang2018spin,webster2018distinct} with antiferromagnetic and non-magnetic ground states do not agree with the experimental results. The ferromagnetic ground state with the out-of-plane moments is also in line with the magnetic measurements \cite{liu2018anisotropic}. Direct comparison of experimental data with the calculated bands \cite{ webster2018distinct} for a monolayer and bulk (FM state) CrI$_3$ is shown in Fig.~\ref{Fig1}(g). It appears that the positions, as well as the shapes of these bands agree better with the calculations for a monolayer CrI$_3$. It is somewhat surprising that our data match moderately well with the calculations for the FM-ground state, even though the experiments were performed far above the Curie temperature, $T_C$. This could indicate that short-range FM fluctuations might already be at play, while the long-range order is not established, as expected for a low-dimensional system \cite{Kosterlitz1973}. By analyzing magnetic entropy data of CrI$_3$, McGuire {\it et al.} \cite{mcguire2015coupling} have also pointed out that the possibility of the presence of short-range magnetic correlations at much higher temperatures than $T_C$. Recently, electron spin resonance (EPR) and magnetic measurements also suggest similar possibilities in other vdW magnetic materials including VI$_3$ \cite{zeisner2020electron,zeisner2019magnetic,jennings1965heat,suzuki2019coulomb,mcguire2017magnetic,tian2019ferromagnetic}. However, we cannot completely rule out the non-magnetic state at high temperature because at this time there are no nonmagnetic calculations that account for SOC and U. Without accounting for these, the calculations result in a metallic ground state, contrasting the observed insulating character of CrI$_3$. Further theoretical and experimental studies are needed for the proper understanding of this issue. We also note that the band structure calculations without SOC do not agree with our results \cite{webster2018distinct,jiang2018spin,zhang2015robust}, indicating that the SOC is essential for the proper description of this system. This is also in line with the observed magnetic anisotropy, as the two-dimensional long-range ferromagnetic order in this system is stabilized by the magnetic anisotropy originating from the SOC \cite{lee2020fundamental,lado2017origin,liu2018anisotropic}.
\begin{figure*}[ht!]
\centering
\includegraphics[width=17cm]{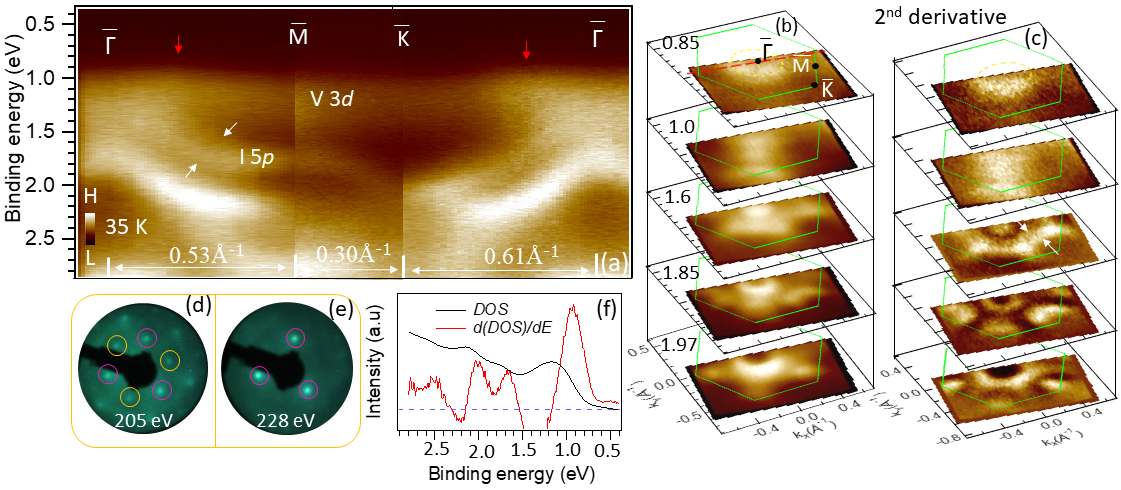}
\caption {Valence band electronic structure of VI$_3$. (a) Probing the band dispersions along $\bar{\Gamma}$-$\bar{M}$-$\bar{K}$-$\bar{\Gamma}$ path using the He I$\alpha$ photons (21.2 eV) at 35 K. Red arrows indicate the position of VBM. White arrows indicate the two bands. (b) ARPES iso-energy contours at different energies, as indicated. (c) Iso-energy contours of intensity\rq{}s second derivative ($d^2I/dE^2$). (d) and (e) LEED patterns showing the three fold symmetry of the surface. The two sets of spots with different intensities are marked by the pink and yellow circles. (f) DOS (black) and its differentiation (red). The blue dotted-line showing the zero intensity reference.}\label{Fig2}
\end{figure*}

We also remark that the calculations based on the hybrid-function (HSE06) place the VBM at ${\Gamma}$, in agreement with our results, but the position of the Cr 3{\it d} bands ({\it w.r.t.} VBM)  is much deeper ($\sim$~3.0~eV) in this calculations \cite{zhang2015robust} than in the experiment ($\sim$~0.85~eV, see Fig.~\ref{Fig1}(e)) or in the GGA  calculations ($\sim$~0.7~eV) \cite{jiang2018spin,zhang2015robust}, LDA+U ($\sim$~0.8~eV) \cite{gudelli2019magnetism}. However, the predicted band gap of $\sim$~1.22-1.28 eV \cite{gudelli2019magnetism,zhang2015robust} is much closer to our results and the optical studies $\sim$~1.24 eV \cite{dillon1965magnetization} than the one obtained from GGA $\sim$~0.7 eV \cite{jiang2018spin,zhang2015robust}. It appears that the HSE06 calculations better predict the band gap, but underestimate the position of the Cr 3{\it d} bands {\it w.r.t.} the VBM.

Now, we turn to VI$_3$. Figure~\ref{Fig2} shows the electronic structure of VI$_3$ along the high symmetry directions of SBZ in the FM phase, at 35~K. The band dispersions along the $\bar{\Gamma}$-$\bar{M}$-$\bar{K}$-$\bar{\Gamma}$ path are shown in Fig.~\ref{Fig2}(a). The iso-energy contours at different energies and their second derivatives are shown in Fig.~\ref{Fig2}(b) and (c), respectively. Figure~\ref{Fig2}(d) and (e) represent the LEED patterns obtained from the VI$_3$ surface at ~200 K. The DOS obtained by integrating the photoemission intensity over the whole Brillouin zone and its derivative are plotted in Fig.~\ref{Fig2}(f). The bands with a predominant V 3{\it d} and I 5{\it p} character are marked in Fig.~\ref{Fig2}(a). Below the V 3{\it d}-dominated flat bands, three more dispersing bands, similar to those observed near the VBM in CrI$_3$, are clearly resolved. The white arrows point to the two nearly degenerate bands marked by I 5{\it p}, seen in both the $\bar{\Gamma}$-$\bar{M}$ and $\bar{\Gamma}$-$\bar{K}$ directions. These two bands are better resolved in the second derivative plots shown in Fig.~\ref{Fig4}(a) and (b). The iso-energy contour of $d^2I/dE^2$ at $\sim$1.6~eV shows these two bands forming two sheets, as indicated in Fig.~\ref{Fig2}(c) by the white arrows. Theoretically, very similar iso-energy features were predicted  to form by Te 5{\it p} states for the prototype system, Cr$_2$Ge$_2$Te$_6$ \cite{li2018electronic}. At higher binding energies, 1.85 and 1.97~eV, these two bands are getting closer and form an electron-like contour with the local minimum at $\bar{K}$. Another hole-like contour centered at $\bar{\Gamma}$ is also identified. Interestingly, all iso-energy contours show a 3-fold symmetry, reflecting the symmetry of the surface. This is also observed in our LEED results. From Fig.~\ref{Fig2}(d) it is clear that two sets of spots (marked with yellow and pink circles) display different intensities. For the electron energy of 228~eV,  only one set of spots are seen (Fig.~\ref{Fig2}(e)), suggesting the trigonal symmetry of the crystal (space group R${\bar3}$) \cite{liu2020critical,kong2019vi3}.

Here, we point out that the recent theoretical studies have demonstrated that AFM ordering in a bilayer CrI$_3$ breaks both time-reversal ($T$) and inversion ($P$) symmetries, leading to a giant nonreciprocal second-harmonic generation \cite{Sun2019} and a magnetic photogalvanic effect \cite{zhang2019switchable}. In the underlying crystallographic structure, the inversion symmetry is preserved, but the AF ordering of the two FM layers within a bilayer breaks it, eliminating the need for the non-centrosymmetric crystal. As a consequence of this symmetry breaking,  the electronic states at $\mathbf{k}$ and $-\mathbf{k}$ in the momentum space are not equivalent. The ARPES would be an ideal experimental probe to directly detect such an effect in an AF bilayer. In the present case, the FM order breaks the time-reversal symmetry, but the inversion symmetry is preserved in bulk in both materials. However, the inversion symmetry is always broken at any surface. Therefore, we might expect that in the FM state, both symmetries are broken in the surface region in both CrI$_3$ and VI$_3$, leading to a situation similar to the one predicted for a CrI$_3$ bilayer. An exciting possibility is that the observed 3-fold symmetry of photoemission intensity in Fig.~\ref{Fig2}(b,c) might be partially caused by $P-T$ symmetry breaking near the surface.
\begin{figure*}[ht!]
\centering
\includegraphics[width=13cm]{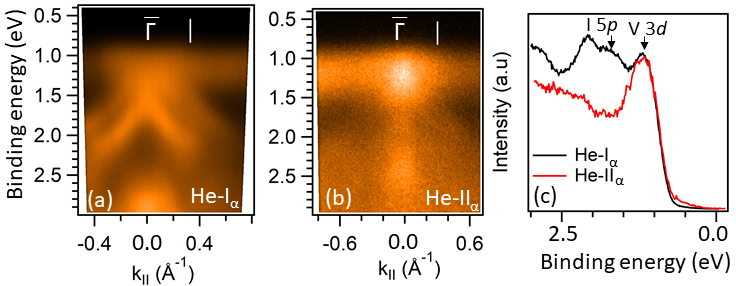}
\caption {Orbital character of the VBM in VI$_3$. (a) and (b) Valence band dispersions around $\bar{\Gamma}$ using He-I$_\alpha$ and He -II$_\alpha$ photons, respectively. (c) EDCs at the momenta indicated by the white vertical lines in (a) and (b). EDCs are normalized to the V 3{\it d} peak.}\label{Fig3}
\end{figure*}

We note that the overall band dispersions in VI$_3$ are in general agreement with the reported calculations \cite{tian2019ferromagnetic} as shown in the supplementary material (Fig. S1). However, as most of the calculations do not account for SOC, a detailed comparison would be somewhat inappropriate. A closer look of the VBM region in Fig.~\ref{Fig2}(a) indicates that the position of the VBM is not at $\bar{\Gamma}$, but at k$_\| \simeq~$0.23~\AA$^{-1}$ (marked by red arrows) in both $\bar{\Gamma}$-$\bar{M}$ and $\bar{\Gamma}$-$\bar{K}$ directions. For better visualization of the VBM position, we show the set of corresponding energy distribution curves (EDC) in the supplementary material (Fig. S2). As a consequence of this displacement, the iso-energy surface at $\sim$ 0.85 eV shows a nearly circular feature around $\bar{\Gamma}$, as can be seen in Fig.~\ref{Fig2}(b) and (c). This is in sharp contrast with the reported theoretical calculations \cite{tian2019ferromagnetic,kong2019vi3}, where VBM is located at $\bar{\Gamma}$. Furthermore, the absence of on-site Coulomb interactions (U) erroneously predicts a metallic ground state \cite{an2019tuning,kong2019vi3}. Our results show a clear band gap, in line with the optical gap of $\sim$ 0.6-0.67 eV observed in this system \cite{kong2019vi3,son2019bulk}. These results clearly indicate the importance of on-site Coulomb interaction for a proper description of this system.

Another important observation contradicting the theoretical reports is that the VBM in VI$_3$ is dominated by the V 3{\it d} character and not by I 5{\it p} as was the case in CrI$_3$ \cite{wang2020electronic,an2019tuning}. A qualitative assessment of the orbital character can be made by comparing the electronic structure of these two materials (Fig.~\ref{Fig1}(a) and Fig.~\ref{Fig2}(a)) where the more localized 3{\it d} orbitals of a transition metal display less dispersion. In the case of  CrI$_3$ these correspond to the features observed around $\sim$ 2.5~eV, whereas in VI$_3$ a flat band at $\sim$ 1.0~eV, most probably corresponds to V 3{\it d} related states. The dominant V 3{\it d} character of the VBM can be also affirmed by comparing the photoemission intensities of the spectra taken with He-I$_\alpha$ and  He-II$_\alpha$. It is well-known that photoemission cross-sections \cite{ yeh1985atomic} of V 3{\it d} states are comparable at these two photon energies, whereas I 5{\it p} states should be an order of magnitude more intense at He-I$_\alpha$ than at He-II$_\alpha$. This would result in a significantly stronger intensity from the V 3{\it d} dominated states in the spectra taken at He-II$_\alpha$. Figure~\ref{Fig3} shows that this is indeed the case. Figure~\ref{Fig3}(a) and (b) are the ARPES spectra around the $\bar{\Gamma}$ point recorded by using He I$_\alpha$ and  He II$_\alpha$ photons, respectively. The corresponding energy distribution curves (EDCs) at the marked momenta are plotted in  Fig.~\ref{Fig3}(c). If normalized to the state closer to the Fermi energy, the spectral intensity of the second peak ($\sim$~1.75 eV) is quite suppressed at He-II$_\alpha$, indicating that the former peak is dominated by V 3{\it d}, while the latter is dominated by I 5{\it p} orbitals. We note that the similar dependence of spectral intensities on the light excitation can be also observed in CrI$_3$, where the I 5{\it p}-derived states are better visible in the He-I$_\alpha$ spectra, Fig.~\ref{Fig1}(f,g). In contrast, the calculations show that the VBM is dominated by iodine states in both materials \cite{wang2020electronic,an2019tuning}.
We note that in some other vanadium based vdW systems, such as VSe$_2$ and VTe$_2$, the less-dispersive V 3{\it d} states are also at the VBM, while the strongly dispersive Se 4{\it p} and Te 5{\it p} states are just below it, similar to our results from VI$_3$ \cite{feng2018electronic,PhysRevB.100.241404}. Similarly, the iodine-derived bands associated with VI$_3$ in the VI$_3$/CrI$_3$ heterostructure \cite{subhan2020large} have their maxima at $\Gamma$ and then disperse downward in both $\Gamma$-$M$ and $\Gamma$-$K$ directions, resembling the states just below the flat band in our experiment. We believe that the observed discrepancy between the current theory and experiment in VI$_3$ could be eliminated by the proper optimization of theoretical parameters, including the exact symmetry of the crystal structure, SOC, U, {\it etc}.

The electronic band gap in VI$_3$ in our study is $\geq$ 0.85 eV (as the position of the Fermi level is $\sim$ 0.85 eV above the VBM) which is higher than the optical band gap, 0.6-0.67 eV  \cite{kong2019vi3,son2019bulk}. The theoretical band gap varies from 0.9 eV for bulk to the 0.97 eV for monolayer \cite{tian2019ferromagnetic}. Our results would place the Fermi level very close to the conduction band minimum (CBM), suggesting a slightly electron-doped system. This may reflect a slight excess of vanadium in the crystal that is thought to partially occupy the vacant interstitial sites within the honeycomb V layers \cite{kong2019vi3} that act as electron donors in this system.

\begin{figure}[ht!]
\centering
\includegraphics[width=8.5cm]{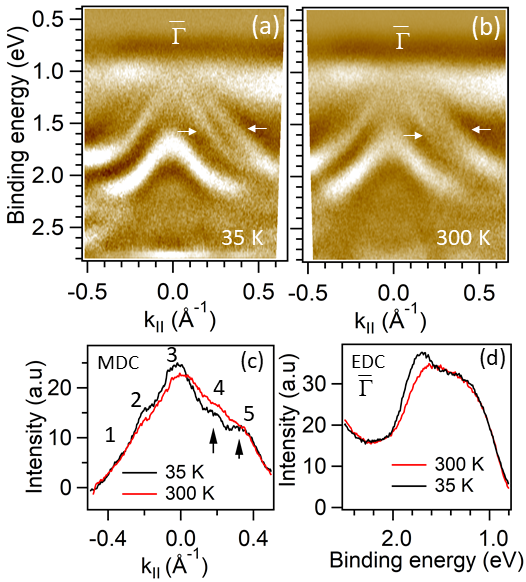}
\caption {Band dispersions below and above magnetic transition temperature in VI$_3$. (a) and (b) the second derivative of ARPES intensity indicating band dispersions along the cut indicated by red dotted line in Fig. 2(b) at 35 K and 300 K, respectively. (c) MDCs at 1.7 eV binding energy at two different temperatures, as indicated. Five peaks can be identified where the peaks 4 and 5 are coming from the two bands indicated by arrows in (a) and (b). (d) EDCs at $\bar{\Gamma}$ at two different temperatures, as indicated.}\label{Fig4}
\end{figure}

Furthermore, to study the effects of structural phase transition and magnetization, we have performed ARPES measurements at room temperature (300 K) and 35 K. Figures~\ref{Fig4}(a) and (b) represent the second derivative of photoemission intensity, indicating band dispersions below and above magnetic transition along the cut indicated by dotted-line (red) in Fig.~\ref{Fig2}(b). Momentum distribution curves (MDC) and EDCs at binding energy of 1.6 eV and at k$_\|$=0 {\AA}$^{-1}$ are presented in Fig.~\ref{Fig4}(c) and (d), respectively. We do not observe major changes in the electronic structure between these two temperatures, except for the fact that the two nearly degenerate bands originating from I 5{\it p} states are better resolved at 35 K than at 300 K and a slight increase in their  separation at low temperature (Fig.~\ref{Fig4}(c) and supplementary material, Fig. S3). In Fig.~\ref{Fig4}(d), the EDCs show that the peak at 1.7 eV also gets sharper and shifts towards the higher binding energy in the FM phase. This is similar to what has been recently observed in another vdW ferromagnet, Cr$_2$Ge$_2$Te$_6$, where the FM transition induces a band-width enhancement of Te 5{\it p} \cite{jiang2020spin} and Cr t$_{2g}$ \cite{PhysRevB.101.205125} related states, along with the energy lowering of 5$p$-e$_g$ hybridized states. We note, however, that relatively minor effects of ferromagnetic transition on the electronic structure of these compounds might not be surprising. Without the applied magnetic field, the magnetic domains should be randomly oriented and their zero net magnetization would not produce a significant effect on the energy bands. As already noted for CrI$_3$, this could also mean that the short-range FM correlations might be present above $T_C$. We also do not detect any change related to the structural transition near 80 K. This is also not surprising as the structural transition involves only the changes in the stacking order along the c-axis while the intra-layer structure remains intact \cite{son2019bulk,tian2019ferromagnetic,an2019tuning}. The weak vdW interaction between the layers and the high surface sensitivity of ARPES would naturally make the changes related to the different stacking of layers unnoticeable.

In summary, we have explored the valence band electronic structure of ferromagnetic vdW materials, CrI$_3$ and VI$_3$ using ARPES and identified the bands with dominant V 3{\it d}, Cr 3{\it d} and I 5{\it p} orbital characters. Our results highlight the importance of correlation effects and SOC in these systems and point out to the drawbacks of some calculation methods. Both the optical and transport measurements \cite{kong2019vi3,wang2018very} show a semiconducting character of these materials, in good agreement with our results. Moreover, our results show the quasi-two dimensional behavior of the electronic structure in these materials. The future experimental studies of the electronic structure of thin flakes or films would offer a better insight into the development of dimensionality in these materials. Also, this would allow to study some truly exotic phenomena, predicted to occur at some thicknesses, including a giant nonreciprocal second-harmonic generation \cite{Sun2019} and the creation of dc current by a linearly polarised light \cite{zhang2019switchable} in an AFM bilayer of CrI$_3$.

\section*{Methods}

The experiments within this study were performed in an experimental facility that integrates oxide-molecular beam epitaxy (OMBE), ARPES, and scanning tunneling microscopy (STM) in a common ultra-high vacuum (UHV) system \cite{Kim2018a}. Bulk CrI$_3$ and VI$_3$ single crystals were grown by the chemical vapor transport method (CVT). The details of sample growth and characterization can be found elsewhere \cite{liu2018three,liu2020critical}. The as-grown samples were clamped to the sample holder under argon atmosphere, cleaved with kapton tape inside UHV, and studied by ARPES. The photoemission experiments were carried out on a Scienta SES-R4000 electron spectrometer with the monochromatized He-I$_\alpha$(21.2 eV) and He-II$_\alpha$(40.8 eV) radiation (VUV-5k). The total instrumental energy resolution was $\sim$ 5 meV and $\sim$ 20 meV at He-I$_\alpha$ and He-II$_\alpha$, respectively. Angular resolution was better than $\sim 0.15^{\circ}$ and $0.4^{\circ}$ along and perpendicular to the slit of the analyzer, respectively. {\it In-situ} low-energy electron diffraction (LEED) measurements were conducted after ARPES measurements to establish the orientation and surface quality. Due to the insulating behavior of the measured samples, the charging was present in the photoemission experiments at the low temperature for VI$_3$ and even at 300 K for CrI$_3$, resulting in energy shifts of photoemission intensity. In the case of VI$_3$ the charging was not present at 300 K, neither at He-I$_\alpha$ nor at He-II$_\alpha$. At 35 K, there was only a $\sim$20 meV shift. Therefore, the VBM recorded at 300 K was taken as a reference and the 35 K data were aligned with that. Similarly, for CrI$_3$, where charging was more severe, the VBM measured at 370 K was taken as a reference and the 300 K data were aligned accordingly. When charging was present, it was steady with time.

\bibliography{sample}

\section*{Acknowledgements}

We thank C. Homes for discussions. This work was supported by the US Department of Energy, office of Basic Energy Sciences, contract no. DE-SC0012704.

\section*{Author contributions statement}

T.V. directed the study and made contributions to development of the OASIS facility used herein. A.K.K. and T.V. designed the study, performed the ARPES and LEED experiments, analyzed and interpreted data and wrote the manuscript. Y.L. and C.P. grew the bulk single-crystals and commented on the manuscript.

\section*{Additional information}
Correspondence and requests for materials should be addressed to A.K.K. (akundu@bnl.gov) or T.V. (valla@bnl.gov).

\section*{Competing interests}
The authors declare no competing interests.

\end{document}